\documentclass{PoS}
\usepackage{color,latexsym,graphics,graphicx,epsfig,amsmath,bbm}

\def\e{{\epsilon} }

\def\P{{\mathcal P}}
\def\V{{\mathcal V}}

\def\S{{\mathcal S}}

\def\jet{{\mbox{\footnotesize jet}}}
\def\hydro{{\mbox{\footnotesize hydro}}}

\def\be{\begin{equation}}
\def\ee{\end{equation}}
\def\bea{\begin{eqnarray}}
\def\eea{\end{eqnarray}}

\def\Eq#1{Eq.~(\ref{#1})}
\def\Eqs#1{Eqs.~(\ref{#1})}
\def\App#1{Appendix~\ref{#1}}
\def\Fig#1{Fig.~\ref{#1}}
\def\Sect#1{Section~\ref{#1}}

\def\Ref#1{Ref.~\cite{#1}}

\title{Jet-medium interaction and the Gubser flow}

\ShortTitle{Jet-medium interaction and the Gubser flow}

\author{\speaker{Li Yan}, %\thanks{A footnote may follow.}, 
        Sangyong Jeon, and Charles Gale\\
        Department of Physics, McGill University, 3600 Rue University, Montr\'eal, QC, H3A 2T8, Canada\\
        E-mail: \email{li.yan@physics.mcgill.ca}}

\abstract{We study the effects of expansion and viscous corrections on the hydrodynamical medium response to a high energy jet parton. More specifically, using a semi-analytical Gubser solution to relativistic fluid dynamics,
modifications to the formed Mach cone, diffusive wake, and the momentum flow of the medium response
along and perpendicular to the jet particle, are analyzed mode-by-mode. This should provide intuition and
guidance for analyses of the experimentally measured jet sub-structure in heavy-ion collisions.}

\FullConference{Critical Point and Onset of Deconfinement - CPOD2017\\
		7-11 August, 2017\\
		The Wang Center, Stony Brook University, Stony Brook, NY}

\begin{document}

\section{Introduction}

One of the interesting problems investigated in heavy-ion experiments carried out
at Relativistic Heavy-Ion Collider (RHIC) and the Large Hadron Collider (LHC) is that of 
jet quenching~\cite{Bjorken:1982tu,Gyulassy:2003mc}, which occurs when %related to the fundamental properties of the strong interaction when
an energetic parton (quark or gluon) goes through dense and hot quark-gluon plasma (QGP). 
Analyses of jet quenching
are expected to reveal  fundamental properties of  the strong interaction, like how an energetic parton deposits energy 
into the medium. In that context, the parameter $\hat q$ 
was proposed and extracted to characterize the averaged transverse momentum broadening
of a jet parton propagating in the medium~\cite{Baier:1996kr}. 
Recently, experiments at the LHC energies have reached an 
unprecedented level of details, in what concerns the analyses of jet sub-structure~\cite{CMS:2014uca}. 
%In additional to the information of jet quenching, 
Those measurements and analyses extend to large jet cone radii~\cite{Chatrchyan:2011sx}, 
where the dominant contributions come from the jet-medium interaction. This therefore presents jet-medium interactions as a class of alternative probe of
information related to the dissipative 
properties of the medium, such as the specific 
shear viscosity $\eta/s$.

A consistent theoretical description of the jet-medium interaction is challenging, 
but several analyses have nevertheless been performed~\cite{CasalderreySolana:2004qm,Chesler:2007sv,Tachibana:2014lja,Qin:2009uh,
Betz:2008ka,Chaudhuri:2005vc,Shuryak:2013cja,Chen:2017zte,Milhano:2017nzm,Floerchinger:2014yqa}.
Part of the complication owes to the fact that the jet-medium interaction involves simultaneously
a hard energy scale from the jet parton, and a soft energy scale
from the evolving background medium. The dynamics of the jet parton normally can be 
captured by perturbative QCD calculations.  The evolution of the
background medium, on the other hand, requires effective modelings such as 
viscous hydrodynamics. It is possible to take into account the effect of 
jet-medium interaction 
in the framework of hydrodynamics, as has been applied in many theoretical
analyses. However, one has to be aware that the evolution of high energy
modes associated with the jet parton may be beyond the validity of hydrodynamics.
The key is to incorporate properly 
a high energy mode in hydrodynamics: a theory in which
the dominant modes are long-wavelength (low energy) excitations. 

In these proceedings, we present a formalism developed in \Ref{Yan:2017rku}, in which
the perturbations induced
by the energy deposition from an energetic parton are decomposed into modes. Then, the 
evolution of these modes can be studied in conjunction with an analytically solved
evolution of a conformal medium, the Gubser flow. 
%The jet-medium interaction can be 
%obtained in the mode-by-mode calculation. 
While long-wavelength modes are safely
compatible with hydrodynamics, short-wavelength modes are damped as a
direct consequence of viscous suppression. This is consistent with the expectation
that modes of longer wavelength are more easily thermalized and absorbed in the
background medium system~\cite{Iancu:2015uja}. The jet-medium interaction perturbs the hydrodynamic fields and leads 
leads to a conical structure. The effects 
of the jet-medium interaction
are then examined in the generated particle spectrum, after summing over modes.

The basic aspects of the theoretical framework are discussed in \Sect{sec:2}, with the equation
of motion describing the jet-medium interaction introduced and solved with respect to
the Gubser flow. The  cone induced from
the jet-medium interaction is described in \Sect{sec:3}. 

\section{Methodology}
\label{sec:2}
We discuss the jet propagation through a medium in terms of the fluid dynamics, and we apply an analytically 
solvable model of the Gubser flow. For simplicity, in the calculations we impose Bjorken boost invariance,
so that the background medium, as well as the jet parton, are realized as boost invariant. This is an idealization, but we shall focus on the features we expect to be robust.  

%Note in this way, 
%the jet parton is \emph{not} realistic, with a knife-shape structure that spans along the space-time
%rapidity. Nevertheless, we would expect the physics conclusion made in this work, will not be
%affected qualitatively. 

\subsection{The jet-medium interaction}

In the framework of hydrodynamics, the system evolution is characterized by a set of equations that reflect conservation laws. 
%of motion, following correspondingly the conservation of energy-momentum, number density, 
%etc. 
>From the conservation of energy-momentum, one has
\be
\label{eq:tmn0}
\partial_\mu T^{\mu\nu} = 0\,,
\ee
where the energy-momentum tensor $T^{\mu\nu}$ is a function of hydrodynamic fields:
Energy density $e$, pressure $\P$ and flow velocity $u^\mu$.
%In the context of jet-medium interaction, 
The  system consists of the background fluid, the jet parton, and their interaction. We thus expect the energy-momentum tensor to be  
\be
\label{eq:tmn}
T^{\mu\nu} = T^{\mu\nu}_\hydro + T^{\mu\nu}_\jet + \delta T^{\mu\nu}\,,
\ee
where the three terms in the right hand side corresponds to the energy-momentum tensor
of the background fluid, the jet parton, and the  jet-medium interaction, respectively. 
%It is a plausible assumption
In this work we shall
treat the contributions from the jet parton to the medium
%the jet-medium interaction, 
as a perturbation, when  
comparing to the 
evolution of the background medium. 
Therefore, in  linearized hydrodynamics %to first order, 
the conservation equation \Eq{eq:tmn0} is separated as
\begin{align}
\label{eq:eom}
\partial_\mu T^{\mu\nu}_\hydro =& 0\,,\cr
\partial_\mu (T^{\mu\nu}_\jet + \delta T^{\mu\nu}) = & 0\,,
\end{align}
where the first equation is the ordinary hydrodynamic equation of motion
describing the evolution of  the background fluid, and the second one focuses on the
effect of jet-medium interaction. Note that at leading order the 
jet-medium interaction does not affect the evolution 
of the background fluid, so that two equations in \Eq{eq:eom} are decoupled.
This procedure can be generalized to higher orders iteratively. The formulation involves the
nonlinear couplings of the perturbations at second order, and is shown in \App{app}. 

The evolution
of the jet parton itself can be rewritten effectively as a source term, $J^\nu=-\partial_\mu T^{\mu\nu}_\jet$.
Accordingly, for the jet-medium interaction one has
\be
\label{eq:jet_equ}
\partial_\mu \delta T^{\mu\nu}=J^\nu\,.
\ee
\Eq{eq:jet_equ} is
the central equation we solve in this work to describe the effects of  jet-medium interaction.
Taking $\delta T^{\mu\nu}$ the linearized perturbations induced by the jet source, 
\Eq{eq:jet_equ} can be expressed explicitly as, 
\begin{subequations}
\label{equ:fluc-eom2}
\begin{align}
\left[\delta w Du_\alpha + w\delta u^\mu d_\mu u_\alpha + (D w  + w\partial\cdot u) \delta u_\alpha 
+ \nabla_\alpha\delta \P + wD\delta u_\alpha + d_\mu\delta \Pi_\alpha^\mu-J_\alpha\right]\Delta^{\nu\alpha}=&0\\
\left[D\delta \e + \delta w\partial\cdot u + d_\mu( w \delta u^\mu)
+w\delta u^\alpha D u_\alpha - u^\alpha d_\mu\delta \Pi^\mu_\alpha
+u^\alpha J_\alpha\right]u^\nu=&0\,,
\end{align}
\end{subequations}
where $w=e+\P$ is the enthalpy density. The perturbations of stress tensor are $\delta \Pi^{\mu\nu}$.

To determine the source term in the jet-medium interaction, we follow a kinetic
approach of the evolution of the jet parton~\cite{Tachibana:2017syd}.
In terms of an effective phase space distribution function $f(t, {\vec x},\omega, {\vec k}_{\perp})$ 
of an energetic jet parton ($\omega$ being energy of the parton), 
the kinetic equation is 
\be
\label{eq:jetf}
\frac{d}{dt} f(t, {\vec x},\omega, {\vec k}_{\perp})=\left[\hat e \frac{\partial }{\partial \omega} + \frac{1}{4}\hat q
\frac{\partial}{\partial {\vec k}_\perp^2} \right ]f(t,{\vec x},\omega, {\vec k}_{\perp})\,,
\ee
where parameters $\hat e$ specifies energy loss, 
and $\hat q$ is the transverse momentum broadening parameter. 
%We neglect the induced jet-medium interaction from the 
%transverse momentum broadening in the evolution of the
%jet source. 
Given \Eq{eq:jetf}, the contribution to the source term in hydro
 from the jet parton energy loss can be found
as
\be
\label{eq:Jmu}
%J^\mu = 
-\int\! \!\frac{d^3{\vec k}}{\omega} k^\mu k^\nu\partial_\nu f
%-\int \!\!d^3{\vec k} k^\mu \frac{d f}{dt}=
% -u^\mu_{jet}\hat e\int\! \!d^3{\vec k} w\frac{\partial f}{\partial \omega}
 =u^\mu_{jet}\hat e\int\! \!d^3{\vec k} f=
 u^\mu_{jet}\hat e n_{jet}(\vec x,t )\,,
\ee
The above derivation is exact regarding a massless jet parton, and the  jet parton velocity 
is $u^\mu_\jet=(1,\vec 1)$. The density of the jet parton $n_\jet$ leads to a space-time
dependence of the source. Especially, it characterizes the trajectory of the jet parton going
through the medium.
%The last two equations come from an integration by parts (assuming jet as massless particles. )
Similarly, the second term related to the momentum broadening leads to
\be
-\frac{\hat q}{4} u^\mu_{jet}\int d^3 k \left(\frac{\partial^2 \omega }{\partial k_{\perp}^{i2}} \right)f(t,{\vec x}, \omega,{\vec k}_{\perp})=
-\frac{\hat q}{4} u^\mu_{jet}\int d^3 k \frac{1}{\omega }\left(2-\frac{k_\perp^2}{\omega^2}\right)f(t, {\vec x}, \omega, {\vec k}_{\perp})
\sim -\frac{\hat q}{2} u^\mu_{jet} n_{jet}/p_{jet}\,.
\ee
The last step is estimated by considering a high energy jet 
particle with only colinear emission: 
$f(t,{\vec x}, \omega,{\vec k}_{\perp})\sim n_{jet}(\vec x,t) \delta(k_\parallel-p_{jet})\delta^2(\vec k_\perp)$.
%thus one may ignore the second term in the expression which is sub-leading by 
%$\theta^2\sim k_\perp^2/k_\parallel^2\sim O(g^2)$. 
We consider the relation $\hat q=4\hat e T$~\cite{Moore:2004tg} 
as a result of fluctuation-dissipation, so that
the momentum broadening contribution is suppressed by a factor of the order of
$O(T/p_{jet})$ comparing
to the source term \Eq{eq:Jmu} induced by jet parton energy loss. Therefore,
we neglect the induced jet-medium interaction from the 
transverse momentum broadening in the evolution of the
jet source, so that 
\be
\label{eq:source}
J^\mu =  u^\mu_{jet}\hat e n_{jet}(\vec x,t )\,.
\ee
Throughout this work, we take into account a $T^2$-dependence of the jet parton energy loss
rate, i.e.,
\be
\hat e = \kappa T^2\,,
\ee
with $\kappa$ a dimensionless constant. We consider two possible scenarios to determine 
$\kappa$. The first concerns a weakly coupled medium system, in which the 
dynamical properties of system can be estimated by a quasiparticle assumption, one
expects $\kappa\approx s/3\eta$~\cite{Majumder:2007zh}. 
Namely, following the dynamics of in a weakly-coupled system, 
the jet parton loses more energy to the 
%weakly coupled 
medium when the medium is less dissipative.
Because
the effect of jet-medium interaction is determined essentially by the amount of energy
deposited from the jet parton, we expect a suppressed jet-medium interaction in a more
dissipative fluid. We refer to this suppression as the \emph{dynamical} viscous suppression.
In the second scenario regarding a strongly coupled medium, there is no obvious relation 
between the coefficient $\kappa$ and $\eta/s$, except a lower bound
to the rate of jet energy loss, $\kappa\approx s/3\eta$~\cite{Majumder:2007zh}. 
Once the jet energy loss rate has
little dependence on the medium dissipative properties, the effect of \emph{dynamical}
viscous suppression is negligible.

\subsection{Gubser flow}

The Gubser flow is an analytical solution of a conformal fluid system, with nontrivial 
expansions along both the radial and the longitudinal directions~\cite{Gubser:2010ui}. 
The analytical solution
of the Gubser flow relies on symmetries: boost invariance in the
longitudinal direction and rotational symmetry in the azimuth. These  conditions
are roughly compatible with the ultra-central nucleus-nucleus collisions in the heavy-ion
experiments, and 
are extremely useful to simplify the solution of hydrodynamic equations of motion. Especially 
in the coordinate system $\bar x^\mu=(\rho, \theta, \phi, \xi)$, 
following the coordinate transformation 
between $\bar x^\mu$ and the Milne space-time, $x^{\mu} =(\tau,r, \phi,\xi)$\footnote{
The proper time is $\tau=\sqrt{t^2-z^2}$, and the space-time rapidity is defined
as $\xi=\frac{1}{2}\log\frac{t+z}{t-z}$.
},
\begin{align}
\label{eq:transf_cor}
\sinh\rho=&-\frac{1-q^2\tau^2+q^2r^2}{2q\tau}\,,\\
\tan\theta=&\frac{2qr}{1+q^2\tau^2-q^2r^2}\,,
\end{align}
one observes the manifest $SO(3)$ symmetry in the $(\theta,\phi)$ coordinates
\be
d s^2  = -d \rho^2 + d\xi^2 + \cosh^2\rho (d\theta^2 + \sin^2\theta d\phi^2)\,.
\ee
The parameter $q$ in the transformation \Eq{eq:transf_cor} 
is a dimensional parameter to be determined  by the system size. The transformation of 
coordinates imply transformations of hydro fields. For instance,
\begin{subequations}
\label{equ:trans-law}
\begin{align}
\epsilon &= \tau^{-4}\bar \epsilon\\
u_\tau &= \tau\left(\frac{\partial\rho}{\partial\tau}\bar u_\rho+\frac{\partial\theta}{\partial\tau}\bar u_\theta\right)\\
u_r&= \tau\left(\frac{\partial\rho}{\partial r}\bar u_\rho+\frac{\partial\theta}{\partial r}\bar u_{\perp}\right)\\
u_{\phi}&= \tau\bar u_{\phi}\\
u_{\xi}&=\tau\bar u_{\xi}\,.
\end{align}
\end{subequations}
Note in the above equations, and in the following,  we use an overbar on a quantity
to denote it being in the $\bar x^\mu$ coordinate system. 
With respect to the $SO(3)$ symmetry 
in $(\theta,\phi)$, and the boost invariance in $\xi$, one is allowed to find the solution of flow
four velocity of the background fluid% in the $\bar x^\mu$ coordinate system as
\be
\bar u^\mu = (1,0,0,0)\,.
\ee
Given the transformations in \Eqs{equ:trans-law}, the background medium flow
four velocity is recovered which gives rise to the Gubser's solution.

The $SO(3)$ symmetry in $(\theta, \phi)$ allows a mode expansion in spherical harmonics. Regarding the perturbations of the
hydro fields, we may write 
\begin{subequations}
\label{eq:modes}
\begin{align}
\delta \bar T =& \bar T\sum_{lm}\int\frac{dk_\xi}{2\pi} t^{lm}(\rho) Y_{lm}(\theta,\phi) e^{ik_\xi \xi}\\
\delta \bar u_i =& \sum_{lm}\int\frac{dk_\xi}{2\pi}\left[v_s^{lm}(\rho)\Psi^{lm}_i(\theta,\phi) 
+ v_v^{lm}(\rho)\Phi_i^{lm}(\theta, \phi)\right]e^{ik_\xi \xi}\\
\delta \bar u_\xi =& \sum_{lm}\int \frac{dk_\xi}{2\pi}v_\xi^{lm}(\rho) Y_{lm}(\theta,\phi)e^{ik_\xi \xi}\,.
\end{align}
\end{subequations}
where $Y_{km}$ is a spherical harmonic. The functions
$\Psi_i^{lm}$ and $\Phi^{lm}_i$ are vector spherical harmonics specifying 
scalar and vector modes in the transverse flow velocity. They satisfy the condition of
a vanishing divergence and a vanishing curl, respectively. Indices $l$ amd $m$ label
harmonic order. The variable $k_\xi$ is a variable conjugate to $\xi$. Once the perturbations
of hydro fields are boost invariant, as what we consider in this work, the summation of modes
depends only on the mode $k_\xi=0$.

%{\color{red}(We need to check the normalization which is affected by the choise of basis.)} 
With the mode decomposition in terms of (vector) spherical harmonics, the equations of motion
for the jet-medium interaction \Eqs{equ:fluc-eom2}
%\Eq{eq:jet_equ} 
can accordingly be written mode-by-mode,
\be
\label{eq:jet_eom}
\frac{\partial\bar \V_{lm}(\rho,k_\xi)}{\partial \rho} = 
-\Gamma_l(\rho,k_\xi)\bar \V_{lm}(\rho,k_\xi) + \bar \S_{lm}(\rho,k_\xi)\,,
\ee
with
\be
\bar \V_{lm}(\rho)=
 \begin{pmatrix}
  t_{lm}(\rho) \\
  v^s_{lm}(\rho) \\
  v^\xi_{lm}(\rho)  \\
  v^v_{lm}(\rho)
 \end{pmatrix}
\ee
$\Gamma_l$ is a matrix whose form was given in \Ref{Gubser:2010ui}. Note that
$\Gamma_l$ does not depend on $m$.
Source $\bar \S$ is given by the mode decomposition of the jet source $\bar J^\mu$ (the 
jet source expressed in the $\bar x^\nu$ coordinate system),
 which leads to
\begin{subequations}
\begin{align}
\bar J_\rho = &\sum_{lm}\int\frac{dk_\xi}{2\pi} c^{lm}_\rho Y_{lm}(\theta,\phi) e^{i k_\xi \xi}\\
\bar J_i = & \sum_{lm}\int\frac{dk_\xi}{2\pi}[c^{lm}_s \Psi_i^{lm}(\theta,\phi) + c^{lm}_v \Phi_i^{lm}(\theta,\phi)]e^{i k_\xi \xi}\\
\bar J_\xi =&\sum_{lm}\int\frac{dk_\xi}{2\pi} c^{lm}_\xi Y_{lm}(\theta,\phi)e^{i k_\xi \xi}\,.
\end{align}
\end{subequations}
and correspondingly,
\be
\bar \S_{lm}(\rho)=
 \begin{pmatrix}
  -\frac{1}{3\bar w}c^\rho_{lm}(\rho) \\
   -\frac{2\bar T \tanh \rho}{3\bar w\bar T'} c^s_{lm}(\rho) \\
  \frac{\bar T}{\bar w(\bar T+H_0 \tanh\rho)}c_\xi^{lm}(\rho)\\
  -\frac{2\bar T \tanh \rho}{3\bar w\bar T'}c_v^{lm}(\rho)
 \end{pmatrix}
\ee
We solve numerically  \Eq{eq:jet_eom} for each mode, and the resulting hydro fields
are obtained after mode summations according to \Eqs{eq:modes}. In practical simulations,
we notice that the evolution of higher order modes receive stronger viscous suppressions,
proportional to $\exp({-l^2\eta/s})$~\cite{Gubser:2010ui}. This sort of 
suppression is purely a consequence of the fluid dynamics. We refer the suppression of 
mode evolution due to viscosity to as the \emph{hydrodynamical} viscous suppression. 
%{\color{red} NB, $\t \S(\rho)$ is complex, since $c_{lm}$'s are complex, unless
%jet is moving along as some particular trajectory so that $\bar\phi=0$.}

\section{Jet-medium interaction in heavy-ion collisions}
\label{sec:3}

\begin{figure}
\begin{center}
\includegraphics[width=0.45\textwidth] {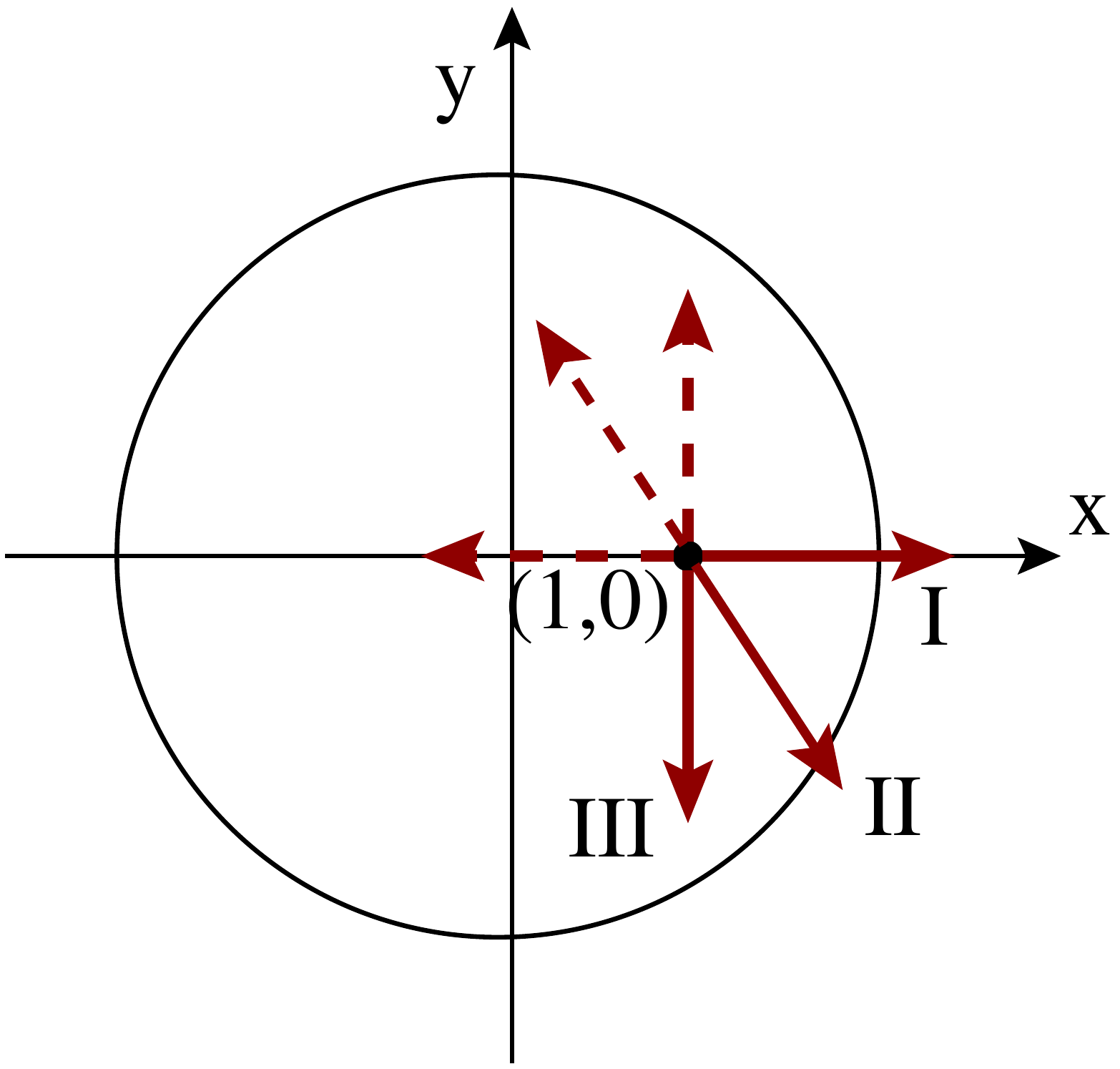}
\caption{
\label{fig:cases}(Color online)
Three events with di-jet considered in this work. Solid 
arrows indicate jet partons generating the near-side (leading) peak
of the observed spectrum, while dashed arrows are
those generating the away-side (sub-leading) peak. 
}
\end{center}
\end{figure}

One may adjust the entropy production of the background fluid system in the 
Gubser flow, according to the experimentally-measured collision events. In this work,
our Gubser flow solution of the background fluid is set to approximate the ultra-central 
Pb+Pb collisions at the LHC $\sqrt{s_{NN}}=2.76$ TeV~\cite{Yan:2015lfa,Shuryak:2013cja}.
On top of the medium expansion, we introduce the source from the external jet 
parton. We consider three representative 
cases of the di-jet partons going through the fluid system, with
their configurations illustrated in \Fig{fig:cases}. In the case I, two back-to-back 
energic partons move along the x-axis, starting from the position $(1,0)$, at a proper time
$\tau_0=1$ fm/c. 
For the case II and III, the pair of partons start from the same position and time, 
but are oriented with an angle of $\pi/3$ and $\pi/2$, respectively.

\begin{figure}
\begin{center}
\includegraphics[width=0.9\textwidth] {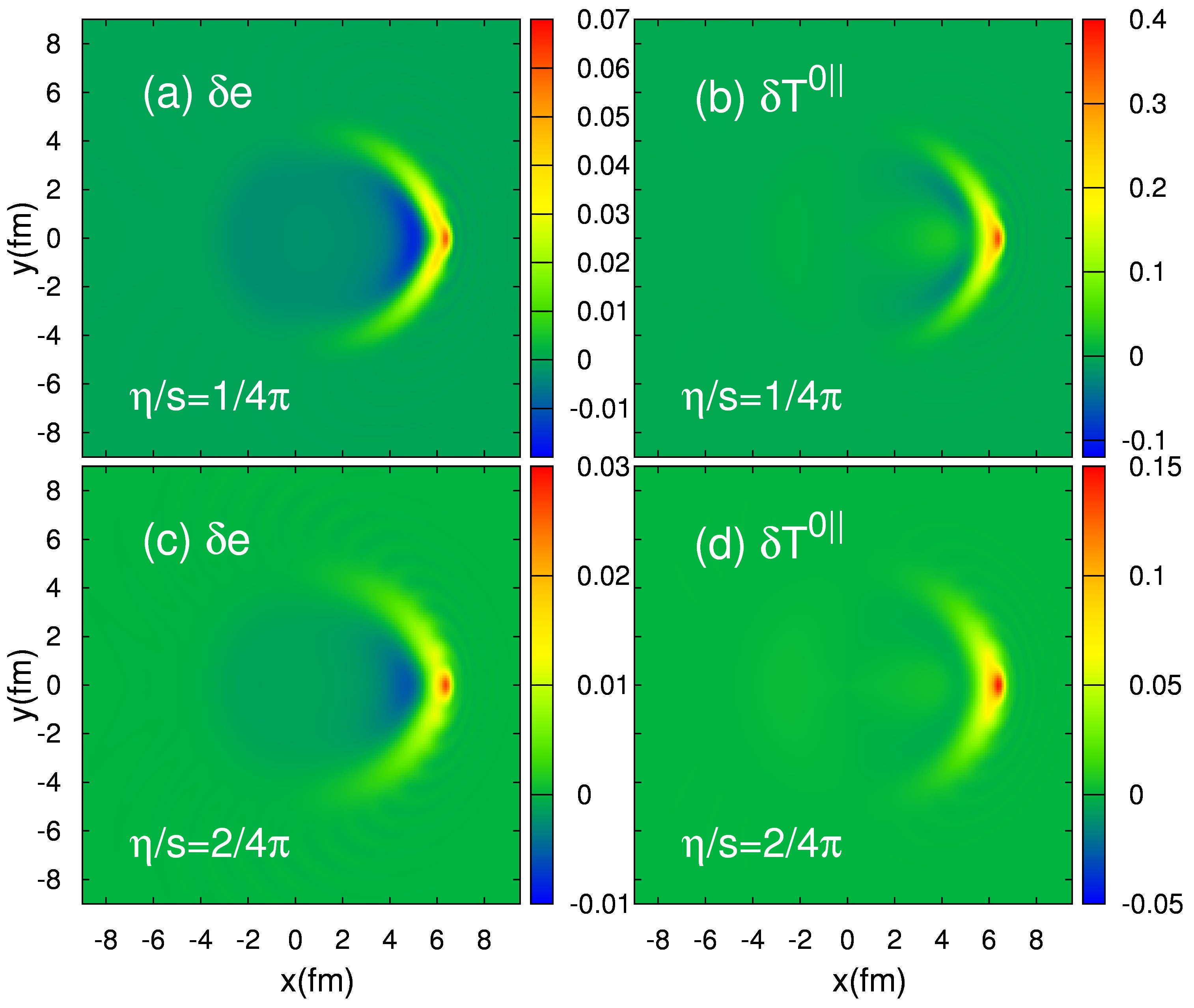}
\caption{
\label{fig:den}(Color online)
%Medium excitations on top of ideal (upper rows)
%and viscous (lower rows) expanding systems with 
%respect to the near-side jet parton of event I, 
%plotted in terms of energy density $\delta e$ (left panels)
%and energy flux %component of energy-momentum tensor 
%$\delta T^{0\parallel}$ (right panels),
%at $\tau=6.0$ fm/c. The colour coding reflects units of GeV/fm$^3$. 
Medium excitations on top of expanding viscous systems with 
$\eta/s=1/4\pi$ (upper row)
and $\eta/s=2/4\pi$ (lower row) with 
respect to the near-side jet parton of event I, 
plotted in terms of energy density $\delta e$ (left panels)
and energy flux %component of energy-momentum tensor 
$\delta T^{0\parallel}$ (right panels),
at $\tau=6.0$ fm/c. The colour coding reflects units of GeV/fm$^3$. 
}
\end{center}
\end{figure}

In the linearized hydro, 
we are allowed to calculate the jet-medium interaction from each parton individually in 
the analysis of di-jet. The resulted medium structure with respect to a di-jet can be 
obtained subsequently
via a linear superposition. To investigate the effect of jet-medium interaction
in the fluid system, we first focus on the jet-medium interaction with respect to
one single jet parton. The results  in \Fig{fig:den} show the  structure induced by the jet-medium interaction associated with the energetic jet parton going outwards in
case I. These figures correspond to the perturbations of energy density $\delta e$ (left 
panel),
and the energy flux along the jet parton $\delta T^{0\parallel}$ (right panel), with
the specific viscosity $\eta/s=1/4\pi$ (upper panel) and $\eta/s=2/4\pi$ (lower panel).

As one expects, the fluid response to a supersonic object results in a conical flow structure, as seen in \Fig{fig:den}. In a static medium the
generated conical flow has a cone angle $\theta^M$ determined by the ratio of the
parton velocity and the speed of sound: 
$\theta^M_0=2\sin^{-1}(c_s/c)\approx 70^o$~\cite{Chesler:2007sv,CasalderreySolana:2004qm}.
The cone, and the cone angle, are both a consequence of coherent 
superposition of the sound wave propagation.
However, with an expanding the background fluid, the propagation of sound modes
is distorted. In the case shown in \Fig{fig:den}, the radial expansion tends to push out
the generated cone structure, hence resulting in a larger cone angle, when comparing with $\theta^M_0$.
Unlike the perturbations of energy density $\delta e$, the energy flux gets additional contributions
from diffusive modes~\cite{Yan:2017rku}. These diffusive 
modes lead to a diffusive wake in $\delta T^{0\parallel}$ behind the shock, which however is not visible in $\delta e$. 

Varying the specific viscosity from $1/4\pi$ to $2/4\pi$ 
in these calculations allows us the examine the effect of 
viscosity on the resulted cone structure from the jet-medium interaction. 
As expected, viscosity suppresses the effect of jet-medium interaction, as seen from the 
smearing of the cone, and a drop in total magnitudes. Regarding the \emph{dynamical}
viscous suppression, we expect an explicit suppression of the magnitude everywhere
by a factor of 2. On the other hand, 
the smearing of the cone is a consequence of the fluid response
to the external parton source, and thus purely the result of \emph{hydrodynamical} viscous 
suppression. Note that in terms of the suppression of magnitudes, \emph{dynamical} 
viscous suppression is the dominant effect.

\begin{figure}
\begin{center}
\includegraphics[width=0.9\textwidth] {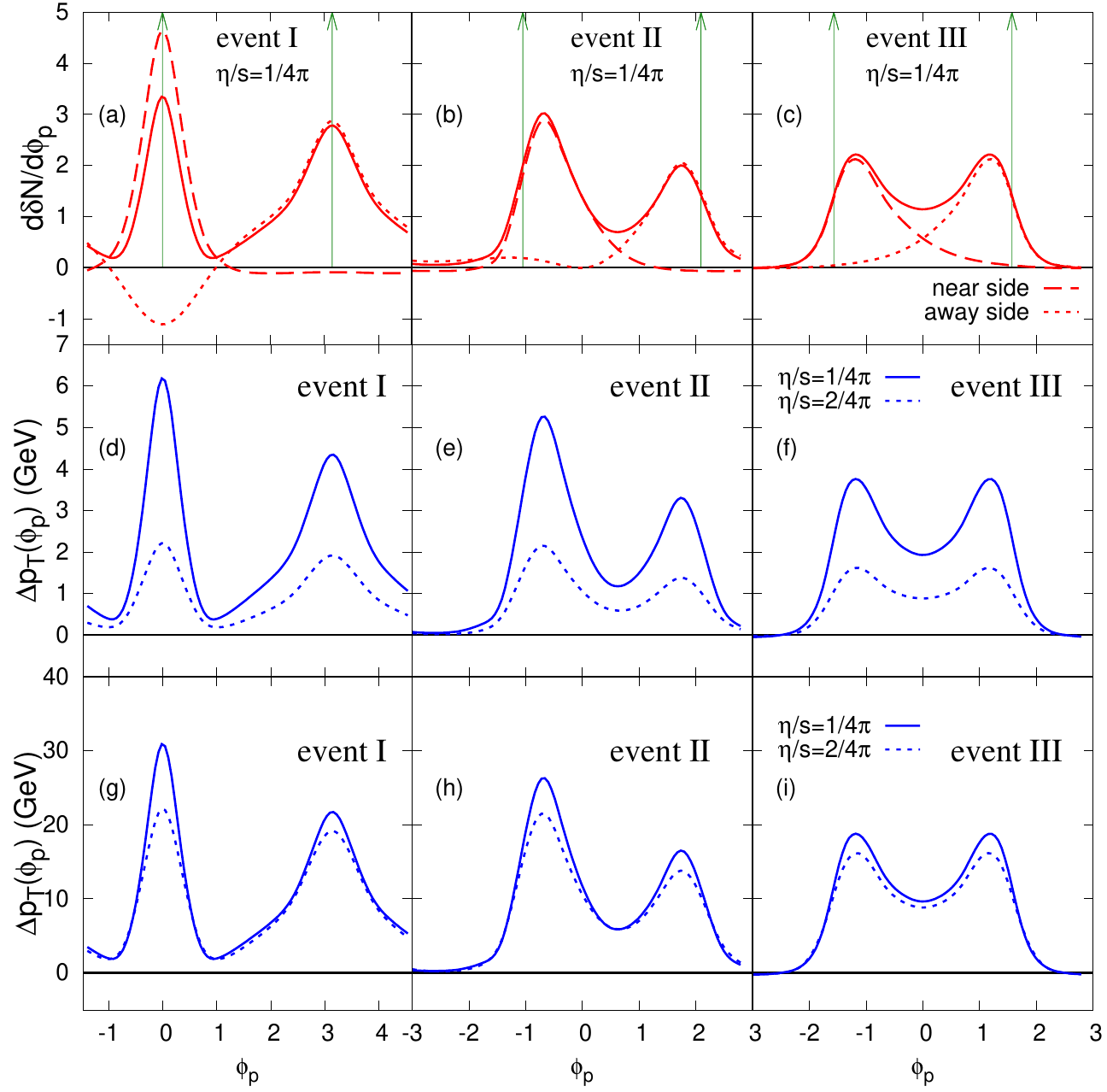}
\caption{
\label{fig:spec} (Color online)
Top row: pion number density generated
from the medium response to jet partons as a function of
azimuthal angle. The near- and away-side number density are shown separately, as well as their sum (solid line). Middle row: transverse energy of the induced
pions from the medium response to a di-jet %jet-medium interaction 
in a weakly-coupled system. 
Bottom  row: transverse energy of the induced
pions from jet-medium interaction for a strongly-coupled 
system. 
See the main text for more details. 
A lower cut of $p_T\ge1$ GeV has been applied.  %taken into account 
%in these plots.
}
\end{center}
\end{figure}

A standard 
Cooper-Frye freeze-out prescription can be applied to the cone structure to convert the medium
response to the jet-medium interaction into the observed particle spectrum. To do so,
we consider a freeze-out of the whole system -- the background and the conical flow -- 
at a proper time $\tau_f=6$ fm/c. The modification to the particle spectrum 
due to the jet-medium interaction is
\be
\label{eq:freeze}
E\frac{d\delta N}{d^3 p}=
\int d \Sigma_\mu p^\mu \delta f\,.
\ee
In \Eq{eq:freeze}, the function $\delta f$ is the difference of the phase-space
distribution at freeze-out, between a system with and without
external jet source. The viscous correction to $\delta f$ is included as well.
For simplicity, we only calculate the spectrum of pion, and ignore any further 
interactions among hadrons.

\Fig{fig:spec} displays the obtained spectrum 
associated with the di-jet in case I (left panel), II (middle panel) and III (right panel).
Although there are two peaks observed in all these cases, associated with the pair of 
jet partons in the di-jet, the background medium expansion distorts differently
the resulted peak structures. Basically, the width of the peak is related to the 
size of the conical structure as a consequence of the Cooper-Frye freeze-out
prescription that converts the fluid cells on the shock structure into particles. 
The larger the cone, the narrower a peak one finds 
in the particle spectrum. For the jet parton going against the medium expansion,
a wider peak is expected, as shown as the dotted lines in \Fig{fig:spec}
(a) and (b). The trajectory of the back-to-back partons are identical in case III, thus
one find the identical two peaks in \Fig{fig:spec} (c). The viscous suppression is revealed
in \Fig{fig:spec} (d), (e) and (f) for the weakly couple system, and \Fig{fig:spec} (g),
(h) and (i) for the strongly coupled system. It should be  emphasized that the difference
between a weekly coupled system and a strongly coupled system, in this work,
only lies in the determination of the rate of energy loss. While for the weakly coupled system
we take the constant $\kappa$ inversely proportional to $\eta/s$, in the strongly 
coupled system $\kappa$ is a constant (somehow large to be compatible
with the physics of strong coupling) independent of $\eta/s$. We note that in a weakly
coupled system, the viscous suppression of the spectrum comes from both the 
\emph{dynamical} viscous suppression and the \emph{hydrodynamical} viscous 
suppression, the viscous effect is strong. However, the dominant effect of viscous
suppression is the \emph{dynamical} viscous suppression. With only \emph{hydrodynamical}
viscous suppression, as in the strongly coupled system, the viscous effect on the
generated particle spectrum is not significant. In fact, it only causes a sizeable reduction 
at the centers of the peaks.

\section{Summary}
In this work, we calculated the jet-medium interaction on an mode-by-mode basis. The
calculation is semi-analytical since the background fluid evolution is analytically solved 
in the Gubser flow. Although it is a simplified analysis owing to the fact that we have taken
into account a boost invariant configuration, including for the jet parton, the
observed effect from viscous suppression is generic. Especially, we realize that there
are two sources of viscous suppression in the obtained conical flow and the 
generated particle spectrum: the \emph{dynamical} viscous suppression and the
\emph{hydrodynamical} viscous suppression. The \emph{dynamical} viscous suppression
accounts for the influence of viscosity on the jet parton energy loss, while
the \emph{hydrodynamical} viscous suppression is purely a consequence of the 
fluid dynamics. Although \emph{hydrodynamical} viscous suppression leads to
smearing the conical flow, and hence a distortion in the observed particle spectrum,
the dominant viscous suppression is the \emph{dynamical} viscous suppression.

%%%%%%%%%%%%%%%%%%%%%%%%
\section*{Acknowledgements}
This work was supported in part by the Natural Sciences and Engineering Research Council of Canada. C. G. gratefully acknowledges support from the Canada Council for the Arts through its Killam Research Fellowship program.

\appendix
\section{Quadratic order hydrodynamics and jet-medium interaction}

In the coupled equations of motion from the conservation of energy-momentum, \Eq{eq:tmn0},
we associate the perturbations
induced by the jet parton with a parameter $\lambda$. The  
%the background fluid and the jet parton the
energy-momentum tensor can be expanded in $\lambda$ as
\begin{align}
&T^{\mu\nu}_\hydro = T^{\mu\nu (0)}_\hydro + \lambda T^{\mu\nu(1)}_\hydro + 
\lambda^2 T^{\mu\nu(2)}_\hydro+
O(\lambda^3)\\
&T^{\mu\nu}_\jet = \lambda T^{\mu\nu(0)}_\jet + \lambda^2 T^{\mu\nu(1)}_\jet
 + O(\lambda^3)
\end{align}
Note that, the above expansion of the energy-momentum tensor of the jet parton 
starts from $O(\lambda)$, because $T^{\mu\nu}_\jet\propto T^2 n_\jet$.
Substituting the above expansions back into \Eq{eq:tmn0} and equating orders,
one has
\begin{align}
\label{eq:geom}
\partial_\mu T^{\mu\nu (0)}_\hydro &= 0\,,\cr
\partial_\mu T^{\mu\nu(1)}_\hydro&= -\partial_\mu T^{\mu\nu(0)}_\jet=J^{\nu(0)}\,, \cr
\partial_\mu T^{\mu\nu(2)}_\hydro&=-\partial_\mu T^{\mu\nu(1)}_\jet=J^{\nu(1)}\,,\cr
&\ldots
\end{align}
Note that the form of $J^{\mu(n)}$ can be determined explicitly according to \Eq{eq:source}.
One recognizes the first two equations at the
linearized order hydrodynamics, as the coupled equations of motion, \Eq{eq:eom}, by identifying $\delta T^{\mu\nu}_\hydro = T^{\mu\nu(1)}_\hydro$. 
By construction, the effective source term $J^{\nu(n)}$, depends on the perturbations of hydro
fields up to $n$-th order. For instance, in the linearized hydro, $J^{\nu(0)}$ depends on the background
hydro fields, $e^{(0)}$, $\P^{(0)}$ etc. When the series of equations are truncated at the next order,
i.e., the quadratic order of hydrodynamics, one needs to solve first the linearized equations for
the linearied perturbations of hydro fields, $e^{(1)}$, $\P^{(1)}$, etc., so as to determine the source
term $J^{\nu(1)}$. Once the quadratic order hydro is solved (the third equation in \Eq{eq:geom}),
the perturbations of hydro fields can be determined as
\begin{align}
&\delta e = e^{(1)} + e^{(2)}\,,\cr
&\delta \P = \P^{(1)} + \P^{(2)}\,,\cr
&\delta u^\mu = u^{\mu(1)}+u^{\mu(2)}\,. 
\end{align}

\label{app}

\bibliographystyle{iopart-num}
\bibliography{refs}

%\begin{thebibliography}{99}
%\bibitem{...}
%...
%\end{thebibliography}

\end{document}